\documentclass{PoS}

\usepackage{epsfig}
\usepackage{graphicx} 
\usepackage[figuresright]{rotating} 
\RequirePackage{xspace}
\usepackage{relsize}
\def\babar{\mbox{\slshape B\kern-0.1em{\smaller A}\kern-0.1em
    B\kern-0.1em{\smaller A\kern-0.2em R}}}

\def\Bz    {\ensuremath{B^0}}
\def\B     {\ensuremath{B}}
\def\Bbar  {\kern 0.18em\overline{\kern -0.18em B}{}}

\def\Bzb   {\ensuremath{\Bbar^0}}
\def\Bz    {\ensuremath{B^0}}
\def\Dm    {\ensuremath{{\rm \Delta}m}} 
\def\Dt    {\ensuremath{{\rm \Delta}t}}

\newcommand{\Btag}{B_{\mathrm{tag}}}

\newcommand{\jprBase}        {Phys.\ Rev.\xspace}
\newcommand{\jplBase}        {Phys.\ Lett.\xspace}
\newcommand{\npBase}         {Nucl.\ Phys.\xspace}
\newcommand{\plb}       [1]  {\jplBase\ B~{\bf #1}}
\newcommand{\jprd}      [1]  {\jprBase\ D~{\bf #1}}
\newcommand{\npb}       [1]  {\npBase\ B~{\bf #1}}

\newcommand{\AmS}{{\protect\the\textfont2
  A\kern-.1667em\lower.5ex\hbox{M}\kern-.125emS}}

\title{How well do we know the Unitarity Triangle? An experimental review} 

\ShortTitle{How well do we know the Unitarity Triangle?}

\author{\speaker{Gabriella Sciolla}\thanks{Representing the \babar\ Collaboration.}\\
        Massachusetts Institute of Technology, 
        Department of Physics,   \\
        Room 26-443,  77 Massachusetts Avenue, Cambridge MA 02139 \\
        E-mail: \email{sciolla@mit.edu}}

\abstract{ 
In the past 10 years our knowledge of the parameters $\rho$ and $\eta$ 
of the Cabibbo-Kobayashi-Maskawa matrix has improved substantially. 
This article reviews the measurements that contributed to this
advance, and discusses their implication 
in terms of understanding $CP$ violation in the Standard Model and beyond.
}

\FullConference{KAON International Conference\\
		 May 21-25 2007\\
		 Laboratori Nazionali di Frascati dell'INFN, Rome, Italy}

\begin{document}

\section{Introduction to the Unitarity Triangle}
According to Kobayashi and Maskawa~\cite{KM}, 
CP violation in the Standard Model (SM) is due to a 
complex phase appearing in the quark mixing matrix,
the Cabibbo-Kobayashi-Maskawa (CKM) matrix. 
Following Wolfenstein's notation~\cite{wolf}, the CKM matrix can be expressed in 
terms of the four real parameters $\lambda$, $A$, $\rho$ and $\eta$ as  
\begin{equation}
V = 
\left(\begin{array}{ccc}
V_{ud} & V_{us} & V_{ub} \\
V_{cd} & V_{cs} & V_{cb} \\
V_{td} & V_{ts} & V_{tb}
\end{array}\right)
= 
\left(\begin{array}{ccc}
1-\lambda^2/2            &   \lambda       & A\lambda^3(\rho-i\eta) \\
-\lambda                 &   1-\lambda^2/2        & A\lambda^2 \\
A\lambda^3(1-\rho-i\eta) &    -A\lambda^2      & 1
\end{array}\right)
+ O(\lambda^4).
\end{equation}

\begin{figure}[htb]
  \label{UT}
  \begin{center}
  \includegraphics[height=0.20\textheight]{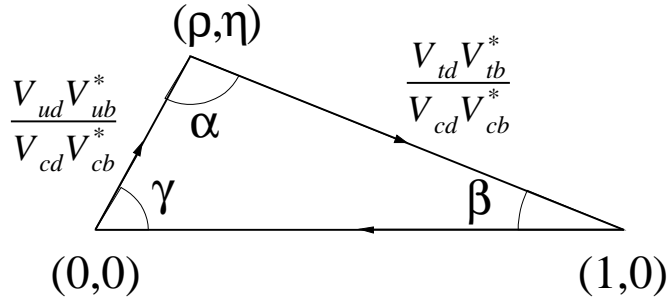}
  \caption{The Unitarity Triangle.}
  \end{center}
\end{figure}

While the parameters  $\lambda$ and $A$ have been precisely known for a 
long time, the  parameters $\rho$ and $\eta$ were poorly measured until recently. 
The parameter $\eta$ is of particular interest, because 
if $\eta=0$ the Standard Model would not be able to explain CP violation. 
If the CKM matrix is unitary, then $V^+V=1$. This implies six unitarity conditions that 
relate the nine elements of the matrix. 
The  condition that relates the first and third columns of the matrix 
can be written as 
\begin{equation}
\frac{V_{ud}V_{ub}^*}{V_{cd}V_{cb}^*} + \frac{V_{td}V_{tb}^*}{V_{cd}V_{cb}^*} +1 = 0.
\end{equation}
This equation represents a triangle in the complex ($\rho,\eta$) plane.
This triangle, knows as the Unitarity Triangle (UT), is depicted in figure~\ref{UT}. 

The study of $B$ and $K$ meson decays allows us to perform a number of measurements that 
set constraints in the ($\rho,\eta$) plane. 
In the Standard Model all measurements must be consistent. 
The presence of New Physics could cause inconsistencies 
for some of the measurements of $\approx 10\%$.
A redundant and precise set of measurements providing constraints in the ($\rho,\eta$) 
plane is therefore essential to test the CKM mechanism and probe for New Physics beyond the Standard Model. 

\section{The measurements} 

The main contributors to this physics program are the two experiments at the asymmetric 
B-factories, \babar~\cite{BaBarDetector} and Belle~\cite{BelleDetector}. 
Collectively, these experiments recorded to date over one billion $B\Bbar$ pairs 
in $e^+e^-$ interactions at the $\Upsilon(4S)$ resonance.  
The large data set and clean experimental environment 
allowed the B factories to measure all sides and angles of the UT\@. 
The two Tevatron experiment, CDF and D0, add important constraints from their measurement of  $B^0_s$ mixing. 
In addition, several kaon experiments (e.g., KTeV, NA48, KLOE) provide complementary information  
by measuring the $CP$-violating parameter $\epsilon_K$ in $K^0$ decays. 

\subsection{CP violation in $B^0$ decays}  
The angles of the UT can be determined 
through the measurement of the time dependent $CP$ 
asymmetry, $A_{CP}(t)$. This quantity is defined as 
\begin{equation}
A_{CP}(t) \equiv \frac{N(\Bzb(t)\to f_{CP}) - N(\Bz(t)\to f_{CP})} {N(\Bzb(t)\to f_{CP}) + N(\Bz(t)\to f_{CP})},  
\label{acpt}
\end{equation}
where $N(\Bzb(t)\to f_{CP})$ is the number of \Bzb\ that decay into the $CP$-eigenstate $f_{CP}$ after a time $t$. 

In general, this asymmetry can be expressed as the sum of two components: 
\begin{equation}
  A_{CP}(t) =  S_f \sin(\Delta m t) - C_f \cos(\Delta m t), 
  \label{acpt2}
\end{equation}
where $\Delta m$ is the difference in mass between $B^0$ mass eigenstates. 
The sine coefficient $S_f$ is related to an angle of the UT, while 
the cosine coefficient $C_f$ measures direct $CP$ violation.

When only one diagram contributes to the final state, the cosine term 
in equation \ref{acpt2} vanishes. 
As an example, for decays such as $B\to\ J/\psi K^0$, $S_f = -\eta_f \times \sin2\beta$, 
where  $\eta_f$ is the  $CP$ eigenvalue of the final state, negative 
for charmonium + $K_S$,  and positive for  charmonium + $K_L$. 
It follows that 
\begin{equation}
A_{CP}(t) = -\eta_f \sin2\beta \sin(\Dm  t),  
\label{acpt5}
\end{equation}
which shows how the angle $\beta$ is 
measured by the amplitude of the time dependent $CP$ asymmetry.

The measurement of $A_{CP}(t)$ utilizes  decays of the $\Upsilon (4S)$ into two neutral $B$ mesons, 
of which one ($B_{CP}$) can be completely  reconstructed  into a $CP$ eigenstate, 
while the decay products of the other ($\Btag$) identify its flavor at decay time. 
The time $t$ between the two $B$ decays is determined by reconstructing the two $B$ decay vertices. 
The $CP$ asymmetry amplitudes are determined from an unbinned maximum likelihood fit 
to the time distributions separately for events tagged as \Bz\ and \Bzb .

\subsection{The angle $\beta$}

\begin{figure*}[th]
\begin{center}
  \includegraphics[height=0.25\textheight]{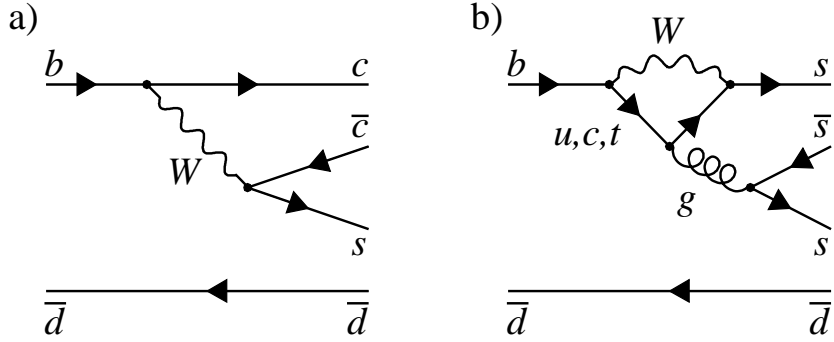}
\end{center}
\caption{Feynman diagrams that mediate the \Bz\ decays used to measure the angle $\beta$: 
a) $\Bz\to\mbox{charmonium}+K^0$; b)  penguin dominated $B$ decays.}  
\label{fey}
\end{figure*}

The most precise measurement of the angle $\beta$ of the UT is obtained in the study of 
the decay $\Bz\to\mbox{charmonium}+K^0$. 	
These decays,
known as ``golden modes,'' 
are dominated by a tree level diagram $b\to c\overline{c}s$ with 
internal $W$ boson emission (figure~\ref{fey}-a). 
The leading penguin diagram contribution to the final state has the same weak phase as the tree diagram, 
and the largest term with different weak phase is a penguin diagram contribution suppressed by $O(\lambda^2)$. 
This makes $C_f=0$ in equation~\ref{acpt2} a very good approximation. 

Besides the theoretical simplicity, these modes also offer experimental advantages because of  their  
relatively large branching fractions ($\sim$$10^{-4}$) 
and the presence of narrow  resonances in the final state,
which provide a powerful rejection of combinatorial background.

The $CP$ eigenstates considered for this analysis are $J/\psi K_S$, $\psi$(2S)$K_S$, 
$\chi_{c1}K_S$, $\eta_cK_S$ and $J/\psi K_L$. 

The asymmetry between the two \Dt\ distributions, clearly visible in figure 3 
is a striking manifestation of $CP$ violation in the $B$ system. 
The same figures also display the corresponding raw $CP$ asymmetry with the 
projection of the unbinned maximum likelihood fit superimposed. 
The measurements from \babar~\cite{sin2bbabar} 
and Belle~\cite{sin2bbelle} are averaged to obtain 
$\sin 2\beta=0.678 \pm 0.026$~\cite{hfag}.  
This measurement  provides the strongest constraints in the $(\rho,\eta)$ plane. 

\begin{figure}[th]
\begin{center}
  \includegraphics[height=0.3\textheight]{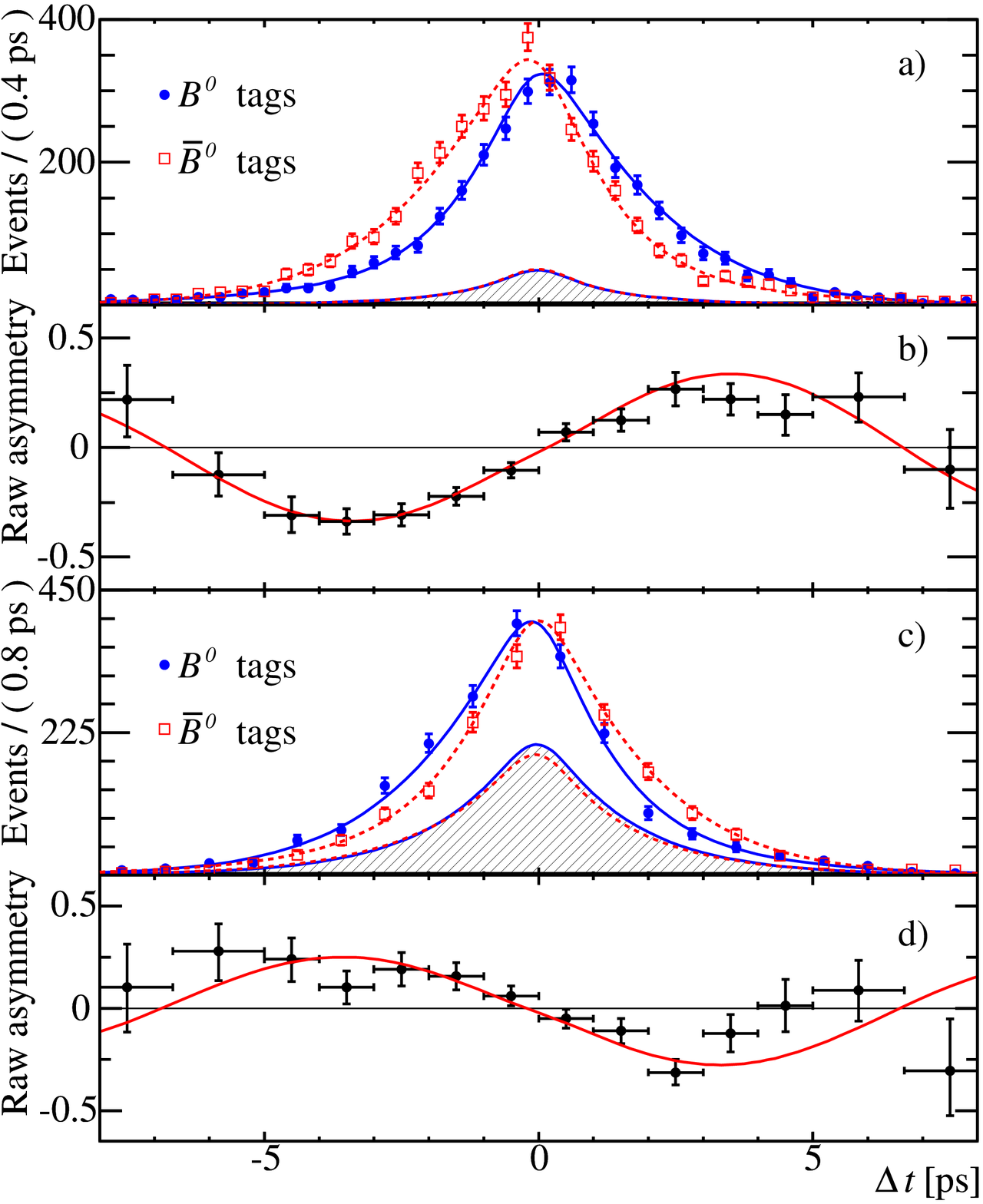}
  \includegraphics[height=0.3\textheight]{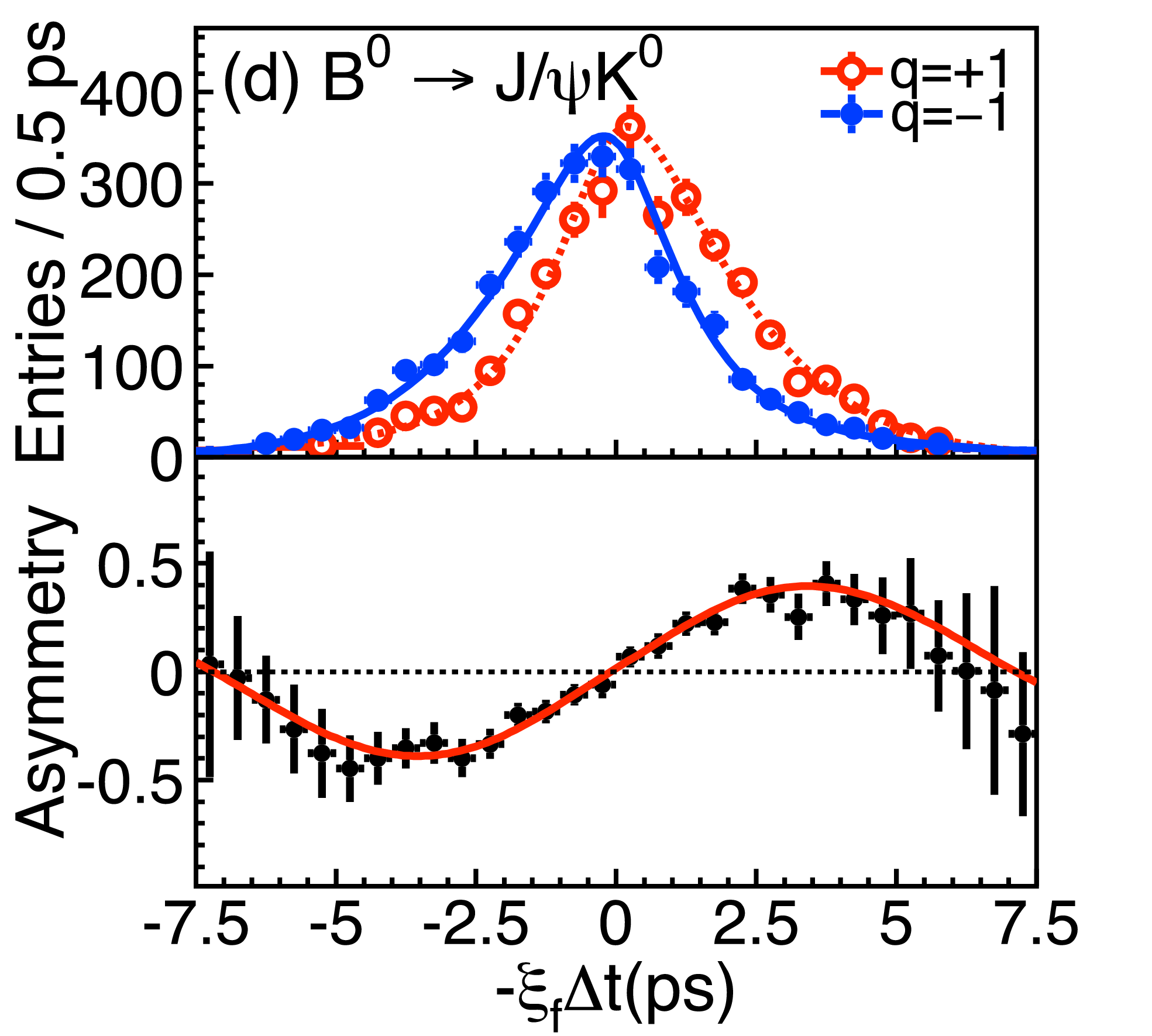}
  \caption{Measurements of $\sin 2\beta$ in the ``golden modes'' by \babar (left) and Belle (right).
    Left plot (\babar): a) time distributions for events tagged as \Bz\ (full dots) or \Bzb\ (open squares) 
    in $CP$ odd  (charmonium $K_S$) final states;  b) 
    corresponding raw $CP$ asymmetry with the projection of the unbinned maximum likelihood fit superimposed;  
    c) and d) corresponding distributions for $CP$ even ($J/\psi K_L$) final states.
    Right plot (Belle): top) time distributions for events tagged as \Bz\ (open dots) or \Bzb\ (openfull dots) 
    in charmonium $K_S$ final states;  b) 
    corresponding raw $CP$ asymmetry with the projection of the unbinned maximum likelihood fit superimposed.}
\end{center}
  \label{sin2bfig}
\end{figure}

An independent measurement of the angle $\beta$ through the study of B decays dominated by  penguin 
 diagrams allows us to search for physics beyond the Standard Model. 
In the SM, final states dominated by $b\to s \overline{s} s $ or $b\to s \overline{d} d $ decays 
offer a clean and independent way of measuring $sin2\beta$~\cite{sPenguin}. 
Examples of these final states are 
$\phi K^0$,  $\eta 'K^0$, $f_0K^0$, $\pi^0 K^0$, $\omega K^0$, $K^+K^-K_S$ and  $K_S K_S K_S$.
These decays are mediated by the gluonic penguin diagram illustrated in figure \ref{fey}-b. 
In presence of physics beyond the Standard Model, new particles such as  
squarks and gluinos, could participate in the loop and affect the time 
dependent asymmetries~\cite{phases}.

\begin{figure}[th]
\begin{center}
\includegraphics[height=0.4\textheight]{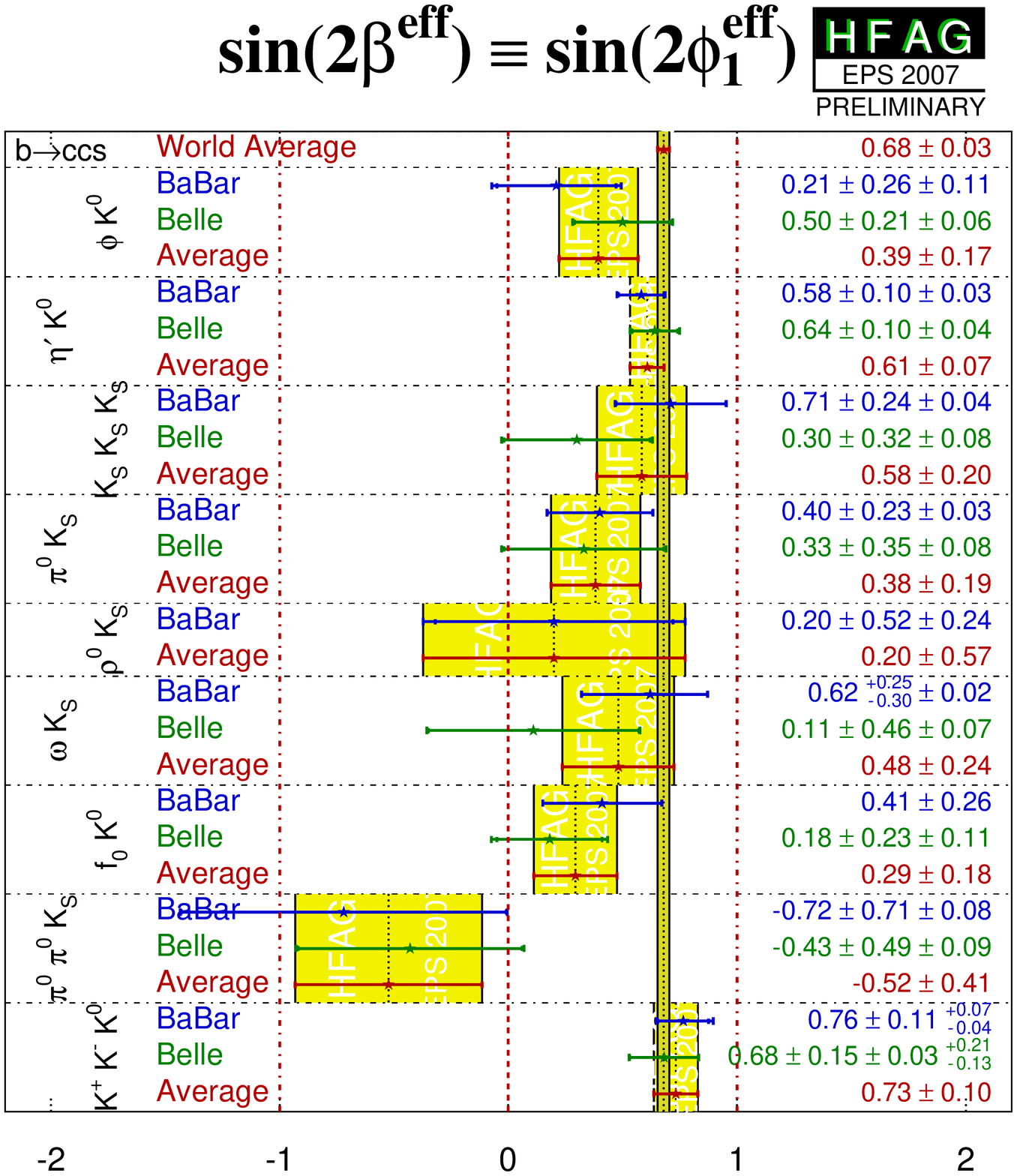}
\end{center}
\vspace{-0.4cm}
\caption{\babar\ and Belle measurements of ``$\sin 2\beta$'' in the penguin dominated  channels. 
The narrow yellow band indicates the world average of the charmonium + $K^0$ final states $\pm 1 \sigma$.}
\label{FigurePenguin}
\end{figure}

A summary of the measurements of $A_{CP}(t)$ in penguin modes
by the \babar ~\cite{sin2b-penguin-Babar}  
and Belle~\cite{sin2bbelle} experiments is reported in figure 
\ref{FigurePenguin}. 
The average of  all the penguin modes, 
$0.56 \pm 0.05$~\cite{hfag}, is about  
2$\sigma$ away from the value of  $\sin 2\beta$ measured in the golden mode. 

\subsection{The angle $\alpha$}

If the decay  $\Bz\to\pi^+\pi^-$ were dominated  by the $b\rightarrow u$ tree level diagram, 
the amplitude of the time-dependent CP asymmetry in this channel would be a clean measurement 
of the parameter $\sin2\alpha$. Unfortunately, the analysis is complicated by 
substantial contributions  from the gluonic  $b\rightarrow d$ penguin amplitudes 
to this final state ($|P/T|\approx 30\%$). 
As a result, in the fit to the time-dependent 
CP asymmetry (equation~\ref{acpt2}) one has to fit for both the sine and the cosine terms. 
The coefficient of the sine term is related to the angle $\alpha$ of the UT through 
isospin asymmetry.  

A similar measurement  can be performed using the decay  $\Bz\to\rho^+\rho^-$.  
This analysis is made difficult by the fact that 
since the $\rho$ is a vector meson, $\rho^+\rho^-$ final states are characterized by three possible angular momentum states, 
and therefore they are expected to be an admixture of $CP=+1$ and $CP=-1$ states. 
However, polarization studies indicate that this final state is almost completely 
longitudinally polarized, and therefore almost a pure $CP=+1$ eigenstate, which simplifies the analysis.  
Additional constraints are obtained by the study of  $B\to\rho\pi$ decays. 

Combining  all \babar\  and Belle results, we measure 
$\alpha =(92^{+10.7}_{-9.3})^{\circ}$~\cite{CKMFitter}. 

\subsection{The angle $\gamma$}
The angle $\gamma$ is measured exploiting the interference between the decays 
$B{}^-\to D^{(*)0}K^{(*)-}$ and $B^-\to\overline{D}{}^{(*)0}K^{(*)-}$, where both 
$D^0$ and $\overline{D}{}^0$ decay to the same final state.
This measurement can be performed in three different ways: 
utilizing decays of $D$ mesons to $CP$ eigenstates~\cite{GWL}, 
utilizing doubly Cabibbo-suppressed decays of the $\overline{D}$ meson~\cite{ADS}, 
and exploiting the interference pattern in the Dalitz plot of $D\to K_S\pi^+\pi^-$ decays~\cite{GGSZ}. 
Currently, the last analysis provides the strongest constraint of the angle $\gamma$. 
Combining all results from \babar\ and Belle, we measure $\gamma=(60^{+38}_{-24})^{\circ}$~\cite{CKMFitter}. 

\subsubsection{ The left side of the Unitarity Triangle}

The left side of the Unitarity Triangle is determined by the ratio of the CKM matrix elements $|V_{ub}|$ and $|V_{cb}|$. 
Both are measured in the study of semi-leptonic $B$ decays. 
The measurement of $|V_{cb}|$ is already very  precise, with errors of the order of 1-2\%~\cite{hfag}.  
The determination of $|V_{ub}|$ is more challenging, mainly due to the large background 
coming from $b\to c\ell\nu$ decays, about 50 times more likely to occur than $b\to u\ell\nu$ transitions. 

Two approaches, inclusive and exclusive, can be used to determine $|V_{ub}|$.
In inclusive analyses of $B\to X_u\ell\nu$,  the $b\to c\ell\nu$ background is suppressed by cutting on a number of kinematical variables. 
This implies that only partial rates can be directly measured, and theoretical assumptions are used to infer the total rate and extract $|V_{ub}|$. 
The theoretical error associated with these measurements is $\approx 8\%$. 
Averaging all inclusive measurements from the \babar, Belle, and CLEO experiments we determine 
$|V_{ub}|=(4.31\pm 0.17 \pm0.35 ) \times 10^{-3}$~\cite{hfag}, where the first error is experimental and the second theoretical. 

In exclusive analyses, $|V_{ub}| $ is extracted from the measurement of the branching fraction $B\to \pi\ell\nu$. These analyses are 
usually characterized by a good signal/background ratio, but lead to measurements with  larger statistical errors due to the the small 
 branching fractions of the mode studied. In addition, the theoretical errors are also larger, 
due to the uncertainties in the form factor calculation. 
Both experimental and theoretical errors are expected to decrease in the future, making this approach  competitive  with the inclusive method. 

\subsubsection{ The right side of the Unitarity Triangle} 

The right  side of the Unitarity Triangle is determined by the ratio of the CKM matrix elements $|V_{td}|$ and $|V_{ts}|$. 
This  ratio can be determined with small ($\approx 4\%$) theoretical uncertainly from the measurement 
of ratio of the $B^0_d$ and $B^0_s$ mixing frequencies. 
The $B^0_d$ mixing parameter $\Delta m_d$ has been measured very precisely by 
many experiments~\cite{hfag}. 
The  $B^0_s$ mixing parameter $\Delta m_s$, which escaped detection for many years 
due to the difficulty in detecting its very fast oscillations, was recently measured 
by the Tevatron experiments~\cite{D0Bsmixing,CDFBsmixing}. 
At the Tevatron, the $B^0_s$ mesons are exclusively reconstructed in their hadronic   or semileptonic  decays. Their 
flavor at production time is inferred by tagging the flavor of the other $B$ hadron produced in the opposite hemisphere, or by looking 
at the sign of fragmentation kaons produced in the same hemisphere.  The time between production and decay of the  $B^0_s$ mesons is 
 then  determined from the measurements of the boost of the  $B_s$  meson and the distance between  
 the interaction point and the $B$ meson decay vertex. 

The value of  $\Delta m_s $ measured by  CDF is $(17.77\pm 0.10 \pm 0.07) \,\mathrm{ps}^{-1}$. 
Combining this measurement with the world average for  $\Delta m_d$,  
one can extract $|V_{td}/V_{ts}|=0.2060\pm 0.0007(\mathrm{exp})^{+0.0081}_{-0.0060}(\mathrm{theo})$. 

An independent determination of  $|V_{td}/V_{ts}|$ can be obtained by the measuring the 
ratio of the branching fractions $BF(\B\to\rho\gamma)/BF(\B\to K^*\gamma)$. 
Recent measurements of the 
branching fractions of  $BF(\B\to\rho\gamma)$ from \babar~\cite{rhogammababar} and Belle~\cite{rhogammabelle} 
yield   $|V_{td}/V_{ts}|=0.201 \pm 0.016(\mathrm{exp}) \pm 0.015 (\mathrm{theo})$. 

The comparison between the two independent  measurements of $|V_{td}/V_{ts}|$, shown in figure 5, 
allows for an independent test of the Standard Model. 

\begin{figure}[htb]
  \label{RhoGamma}
  \begin{center}
  \includegraphics[height=0.40\textheight]{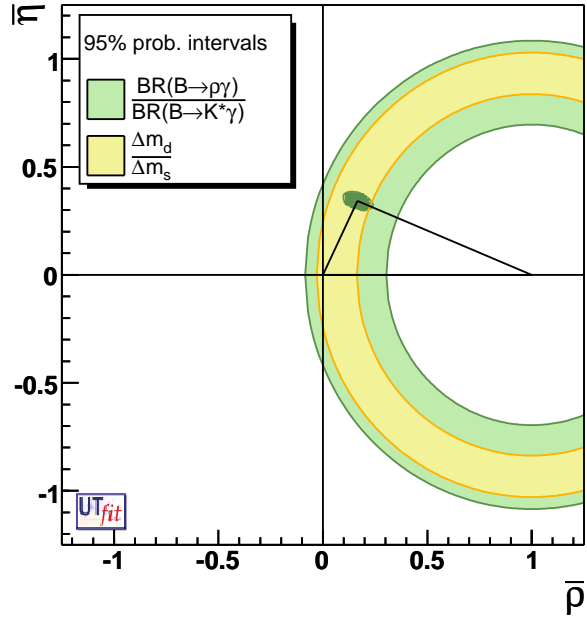}
  \caption{Comparison between the measurements of the right side of the UT from $B^0$ mixing (yellow band) and 
           from $BF(\B\to\rho\gamma)/BF(\B\to K^*\gamma)$ (green band) from reference~\cite{utfit}.}
  \end{center}
\end{figure}

\subsubsection{ Measurement of $\epsilon_K$ }  

Kaon physics contributed the first constraint in the $(\rho , \eta )$ plane 
with the measurement of 
the parameter $\epsilon_K$ in the study of $CP$ violation in neutral kaon decays.  
The parameter $\epsilon_K$ is defined as 

\begin{equation}
  \epsilon_K \approx \frac{2}{3} |\eta_{+-}| + \frac{1}{3}  |\eta_{00}|
  \label{epsilonK}
\end{equation}

where 
\begin{equation}
   |\eta_{+-}| =  \frac{A(K_L\rightarrow \pi^+\pi^-)}{A(K_S\rightarrow \pi^+\pi^-)}  =
   \sqrt{ \frac{BF(K_L\rightarrow \pi^+\pi^-)}{\tau_{K_L}}  \frac{\tau_{K_S}}{BF(K_S\rightarrow \pi^+\pi^-)} } 
\end{equation}
and 
\begin{equation}
   |\eta_{00}| =  \frac{A(K_L\rightarrow \pi^0\pi^0)}{A(K_S\rightarrow \pi^0\pi^0)}  =
   \sqrt{ \frac{BF(K_L\rightarrow \pi^0\pi^0)}{\tau_{K_L}}  \frac{\tau_{K_S}}{BF(K_S\rightarrow \pi^0\pi^0)} }. 
\end{equation}

The 2006 PDG world average, 
including  the latest measurements of the neutral kaon branching fractions from KTeV, NA48 and KLOE,
is $|\epsilon_K|=(2.232\pm0.007)\times 10^{-3}$~\cite{PDG06}. While the experimental 
accuracy on this measurement is remarkable (0.3\%), the corresponding constraints
on the ($\rho , \eta$) plane are not very stringent, 
mainly due to the theoretical uncertainties in the calculation of the bag 
parameter $B_K$ from Lattice QCD. 

\begin{figure}[h]
\begin{center}
 \includegraphics[height=0.5\textheight]{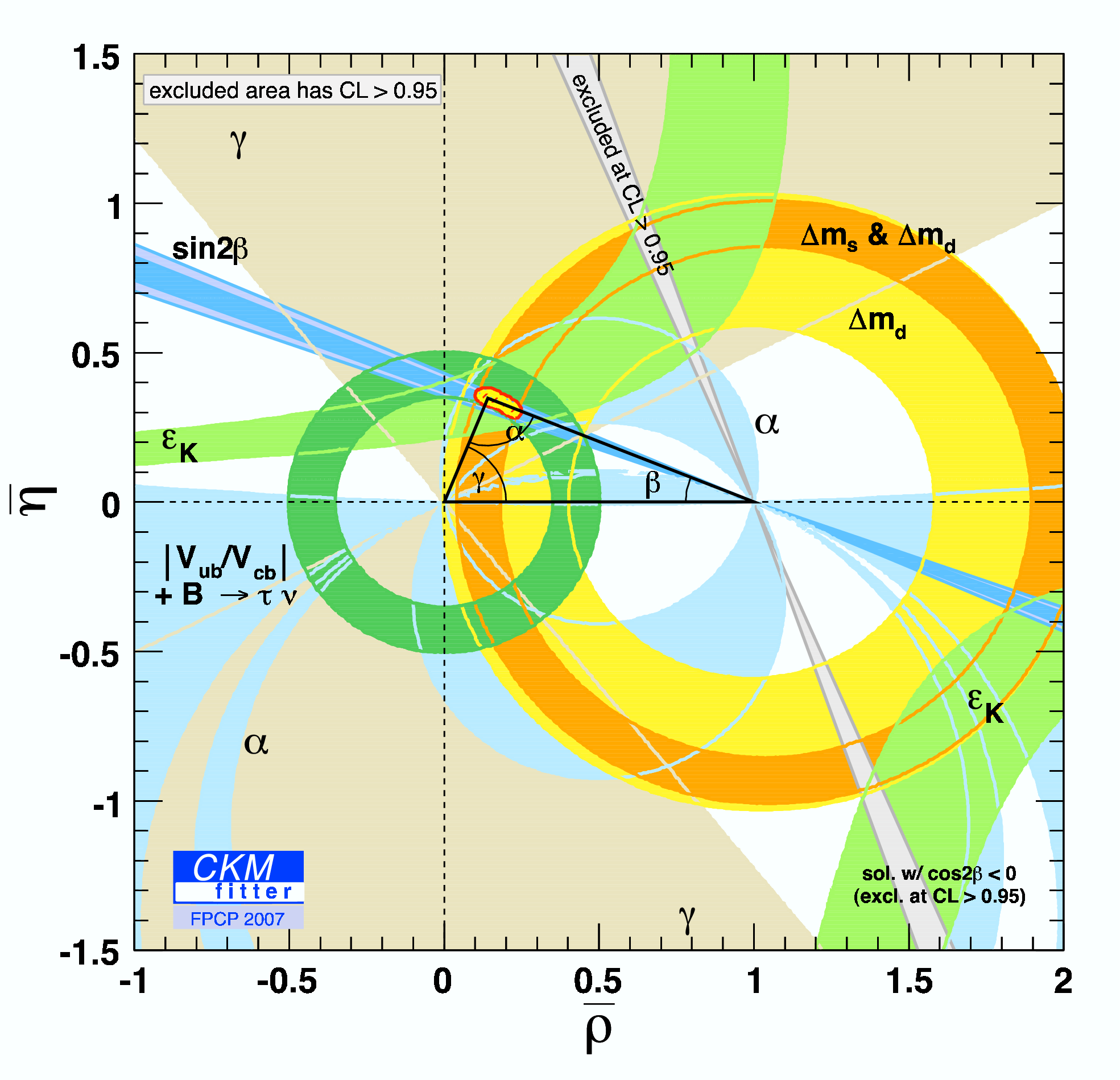}
 \caption{  Constraints on the apex of the Unitarity Triangle resulting from all measurements.        }  
 \label{rhoeta}
\end{center}
\end{figure}

\section{ SUMMARY AND CONCLUSION  }

Precise and redundant measurements of the sides and angles of the 
Unitarity Triangles 
provide a crucial test of  $CP$ violation in the Standard Model. 
The constraints on the ($\rho ,\eta$) plane  due to the 
measurements described in this article are illustrated in figure~\ref{rhoeta}. 
The comparison shows good agreement between all  
measurements, as predicted by the  CKM mechanism.

The accuracy of the measurements is now approaching a few percent. 
This is the level of precision needed for detecting O(0.1) effects
expected from New Physics.
Additional data  from the B factories, 
results from new-generation flavor experiments, and  progress 
in theory especially lattice QCD, will be key to observing 
physics beyond the Standard Model in the flavor sector.

\end{document}